# APPLYING GENERATIVE ADVERSARIAL NETWORKS TO

# INTELLIGENT SUBSURFACE IMAGING

# AND IDENTIFICATION

By

William Rice

Dr. Dalei Wu
Professor of Computer Science
(Chair)

Dr. Yu Liang
Professor of Computer Science
(Co-Chair)

Dr. Li Yang
Professor of Computer Science
(Committee Member)

APPLYING GENERATIVE ADVERSARIAL NETWORKS TO

INTELLIGENT SUBSURFACE IMAGING

AND IDENTIFICATION

By

William Rice

A Thesis Submitted to the Faculty of the University of
Tennessee at Chattanooga in Partial
Fulfillment of the Requirements of the Degree
of Master of Science: Computer Science

The University of Tennessee at Chattanooga
Chattanooga, Tennessee

May 2019

ABSTRACT

To augment training data for machine learning models in Ground Penetrating Radar (GPR) data classification and identification, this thesis focuses on the generation of realistic GPR data using Generative Adversarial Networks. An innovative GAN architecture is proposed for generating GPR B-scans, which is, to the author's knowledge, the first successful application of GAN to GPR B-scans. As one of the major contributions, a novel loss function is formulated by merging frequency domain with time domain features. To test the efficacy of generated B-scans, a real time object classifier is proposed to measure the performance gain derived from augmented B-Scan images. The numerical experiment illustrated that, based on the augmented training data, the proposed GAN architecture demonstrated a significant increase (from 82% to 98%) in the accuracy of the object classifier.

ACKNOWLEDGEMENTS

To begin, I would like to acknowledge my wife, Shannon, she was here for all of the ups, downs, and how every new project is "The hardest problem I've ever solved". I could not have made it this far without her by my side. Furthermore, I would like to thank Dr. Wu, Dr. Liang, and everyone in the lab for giving me the opportunity to conduct research in an academic environment and facilitate my growth as a researcher. In addition, I would like to thank Mike Tatum, Brent Spell, Josh Ziegler, and the rest of the team at Pylon ai. From the very beginning they challenged me, which enabled me to grow in the field of machine learning. Without them, I would not have been able to undertake a research project as challenging as this. I also would like to include a special thank you to my thesis committee. The road to defense has been rocky, but we finally made it.

TABLE OF CONTENTS

## LIST OF TABLES

LIST OF FIGURES

# LIST OF ABBREVIATIONS

GPR, Ground Penetrating Radar

GAN, Generative Adversarial Network

GANs, Generative Adversarial Networks

ANN, Artificial Neural Network

CNN, Convolution Neural Network

YOLO, You Only Look Once

RCNN, Region Convolution Neural Network

EM, Electromagnetic

RNN, Recurrent Neural Network

NN, Neural Network

FDTD, Finite Difference Time Domain

TP, True Positive

FP, False Positive

TN, True Negative

FN, False Negative

PVC, Polyvinylchloride

STFT, Short Term Fourier Transform

MSE, Mean Squared Error

KLD, Kullback-Leiber divergence

JSD, Jensen-Shannon divergence

## LIST OF SYMBOLS

$\Sigma$, Summation

$P$, Probability of event

$f$, Function that accepts arguments

$\lambda$, Penalty term for loss function

$KL(P||Q)$, KL divergence of P and Q

$W(p_r, p_\theta)$, Wasserstein distance between the prior and posterior distribution

$\alpha$, learning rate

$\mathbb{E}$, Expected

$\hat{\theta}$, output vector

sup, suprema

$\max_{w \in W}$, maximum of w in W

$\mathbb{E}_{x \sim p_r}[f_w(x)]$, expected distribution of function when input is x

$\lambda(\|\nabla_{\hat{x}} D_w(\hat{x})\|_2 - 1)^2$, gradient penalty term

$\phi$, frequency transformation

$L^{(i)}$, Loss at iteration i

$z$, latent distribution

$x \sim \mathbb{P}_r$, sample x that has probability distribution $\mathbb{P}_r$

$D_w$, discriminator weight

$\hat{y}$, predicted value

$L_2^{(i)}$, second loss term at iteration i

$G(z)$, generator output

$\nabla$, gradient

$n$, batch size

# CHAPTER 1

# INTRODUCTION

## 1.1 Problem Statement

Model performance in machine learning is heavily dependent upon the availability of training data. Furthermore, in situ applications of machine learning require data to be indicative of the real world environment the applications plan to infer in. For buried object detection, this would be a B-scan image that bears a close resemblance to B-scans collected in the field. However, these data are not widely available, and when available, the number of samples is few. In situations like this, data augmentation can produce many samples to enhance classification of buried objects [25]. In the world of Ground Penetrating Radar, an open source software named *gprMax* is used to simulate the presence of underground objects [30]. This software is based on Finite Difference Time Domain (FDTD) [17], a numerical method to solve Maxwell's equations that govern waves propagation within a specific medium. The problem with this type of simulated data is that it bears little resemblance to a B-scan that would be obtained in the real world. Furthermore, due to the complexity of FDTD, time to complete a single simulation can take several hours. Thus making the synthesis of a large set of images for training data, virtually impossible. To solve this problem, we propose a novel generative model architecture to synthesize realistic B-scans in real time. In addition, to benchmark the generative model, a classifier is produced that is capable of running in real time on an edge server.

## 1.2 Significance of the Study

The significance of this study is to demonstrate that generative models can be successfully applied to GPR data. Furthermore, the generated data can be used in lieu of field collected samples to improve classifiers for the detection of underground objects.

## 1.3 Objectives of the Study

The objectives of this study are three fold. The first is to establish a generative architecture for the generation of pseudo realistic B-scans. The second is to develop a real time classifier, capable of being deployed on an edge sever. The final is to incorporate frequency information into time-domain architectures.

## 1.4 Proposed Approach

In this thesis, Generative Adversarial Networks (GANs) will be investigated for GPR data augmentation for object identification. Both simulated and real GPR data will be considered as inputs to the generator of a GAN to generate realistic GPR data. Based on the feature analysis of GPR data, a novel objective function and the architecture of GAN are proposed. An algorithm for GAN training with different types of training data is developed. The impact of GAN-synthesized data on the performance of GPR image classification is evaluated. To the best knowledge of the authors, very few work has been done on studying GANs for GPR data analysis in a united framework combining data augmentation and data classification. A detailed diagram of the overall system structure is depicted in Figure 1.1

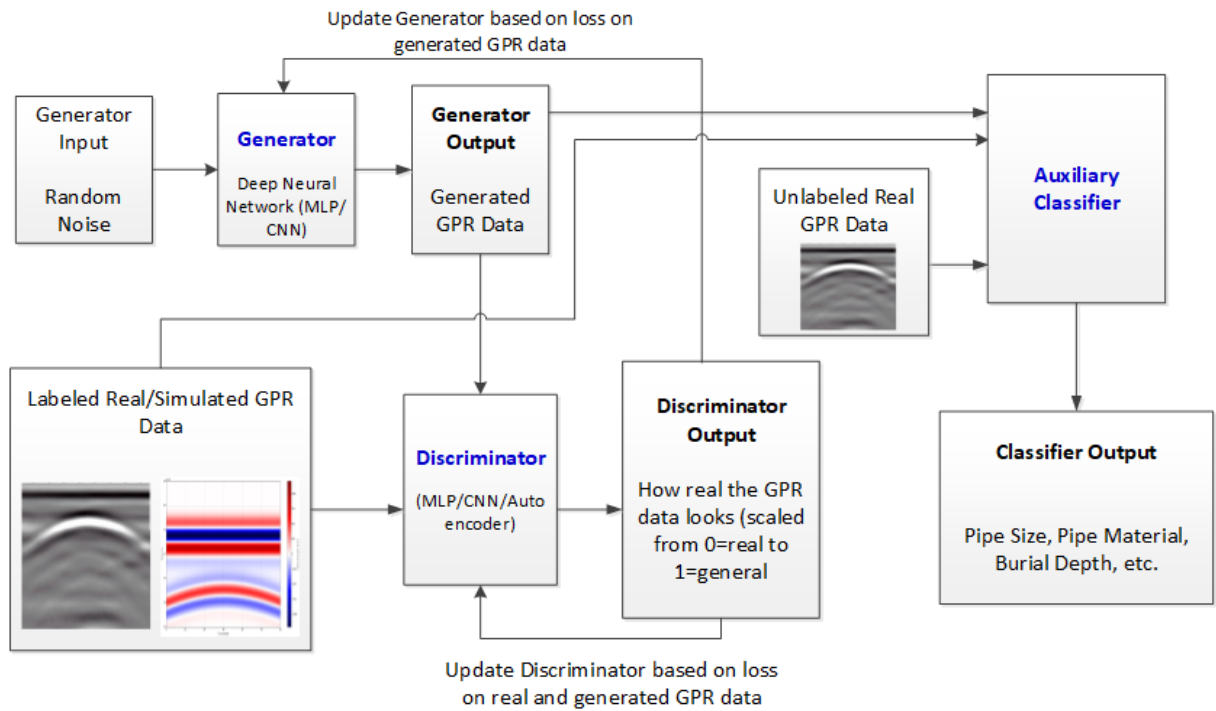

Figure 1.1

Proposed GAN Architecture

## 1.5 Organization of Thesis

The remainder of the thesis is organized as follows: Chapter 2 begins with an introduction to theoretical background for the information contained herein. Chapter 3 is the proposed methodology of the experiments is established along with the architectures of the proposed models. Chapter 4 is the discussion of the results. In Chapter 5, the thesis is concluded with final thoughts and an outline for future work.

CHAPTER 2

BACKGROUND

## 2.1 Introduction

This chapter begins with the theory and applications of Ground Penetrating radar. Then, the necessary background and theory behind Neural Networks is briefly presented. Next, information is then extended with an in depth look at the principle and applications of generative models, which presents the basis for an introduction to Generative Adversarial networks where a large part of this thesis is contained. In support of this, the traditional means of evaluating generative models is explored. Finally, the principles of object identification and classification are presented to introduce the real time classification model.

## 2.2 Ground Penetrating Radar

Ground Penetrating Radar (GPR) is one of the most widely used non-destructive techniques for subsurface imaging and detection of underground objects, such as landmines, utilities, and archaeological artifacts. In the GPR scanning process, an electromagnetic wave is propagated into the target subsurface medium through a transmitting antenna, and upon reflection of the underground object, returned to a receiving antenna. This process is carried out across the above-ground surface for multiple passes. As shown in Figure 4.3 (a), a reflection signal scattering from the underground object can be received and recorded by the receive antenna. As the GPR antennas moves, the reflection signals from the buried object from a number of different angles are received with varying wave propagation distance and amplitude. As a result, the buried object exhibits the hyperbolic image feature in the radar image in Figure 4.3 (b). For each hyperbola composition point, its time index and signal amplitude are determined by a number of object properties, including the depth, size, material type and object shape and the dielectric constant of burying medium. Each time the signal is sent into the material the reflected signal is measured as an A-scan signal, which is a 1-dimensional representation of the signal. A series of A-scans, when concatenated sequentially, form a 2-dimensional high resolution image called B-scan, as shown in figure 2.1. Within such B-scan images, underground objects appear particularly as hyperbolic-shaped signatures. The detection of buried objects can be therefore considered as the detection of reflected hyperbolas in GPR images.

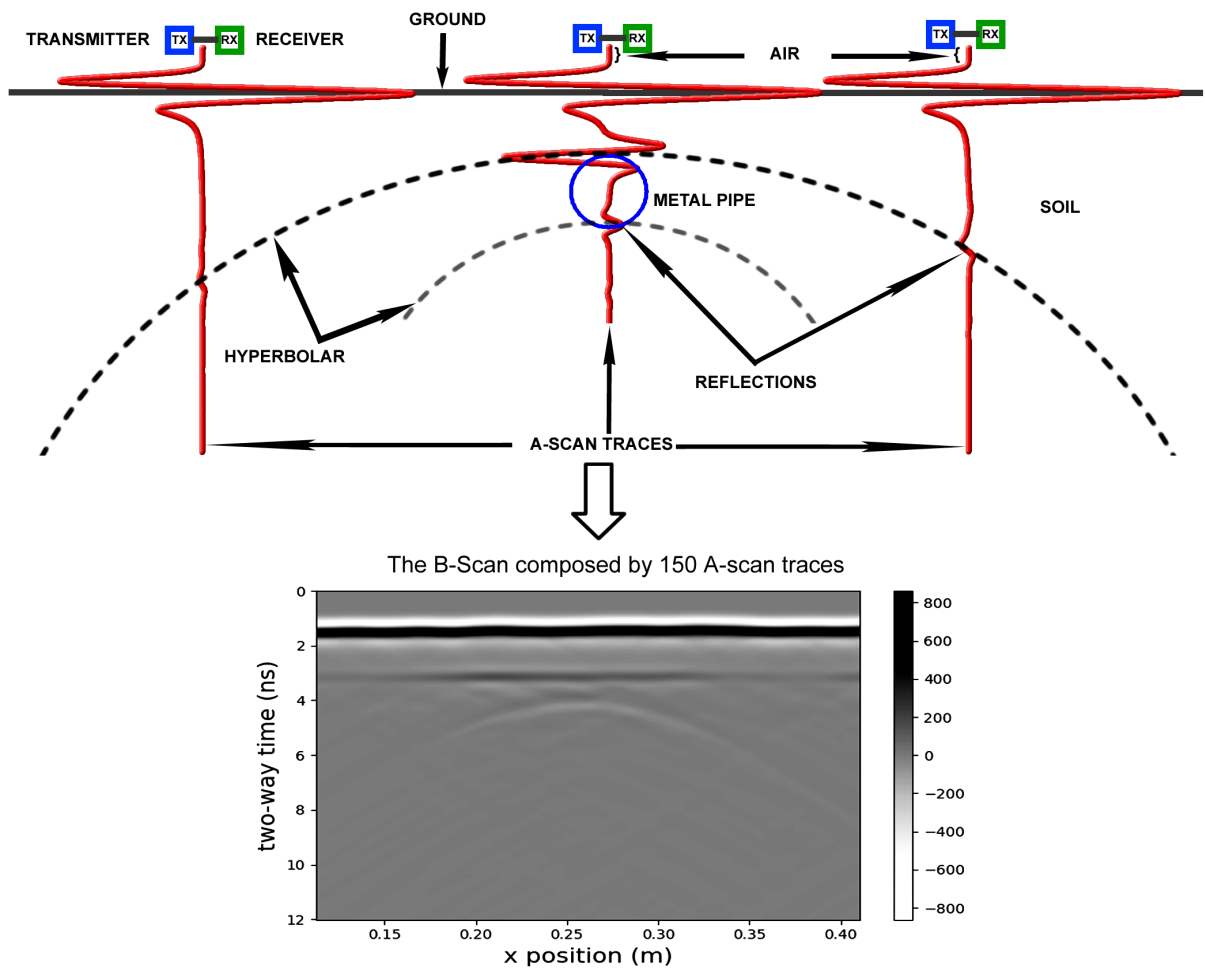

Figure 2.1

Generator and Discriminator Architecture

### 2.2.1 Hands-on Collection

In March 2019, our team was able to walk alongside University of Vermont and get hands on experience in the collection of GPR data. Figure 2.2 depicts a real B-scan in the location that it was taken. The collection process took several days to complete, and from this process we were able to obtained around 20 images. As one can see, collection in the field is a time consuming process. The collection time coupled with the amount of scans collected, is the primary reason GPR data are few.

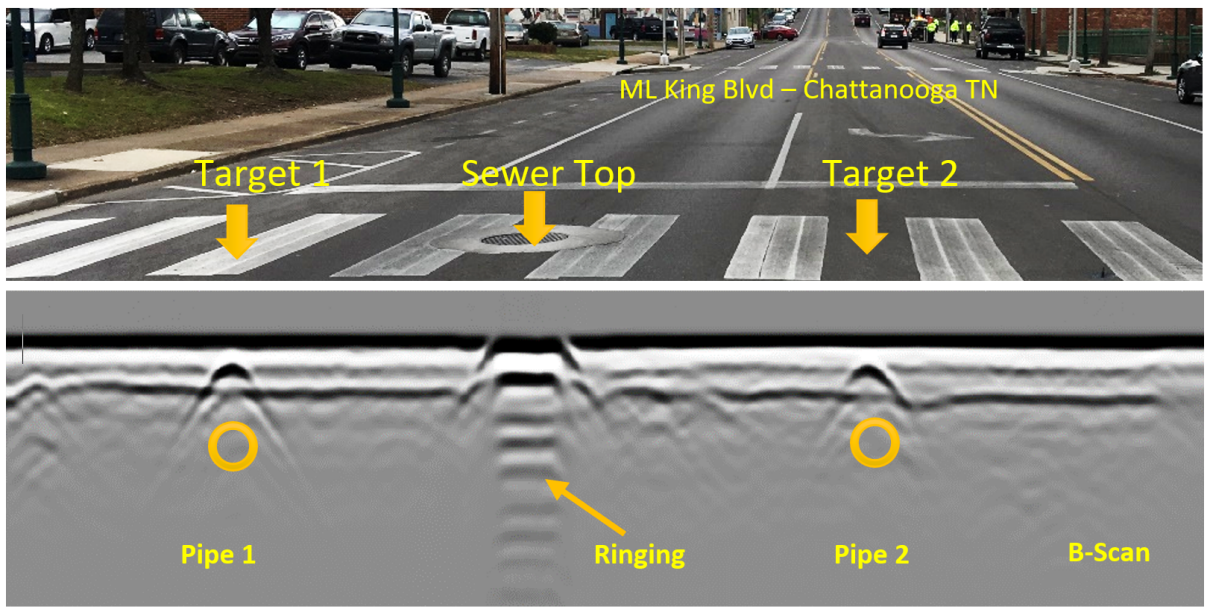

Figure 2.2

GPR Data Collection on MLK

### 2.2.2 B-scan Feature Processing

B-scans, along with other features, are commonly analyzed to detect or identify subsurface objects. As an alternative to visual examination of B-scans by GPR technicians, machine learning techniques have been applied to analyze B-scans for object detection [34, 29, 7, 25, 19]. By combining Hilbert transform and classic artificial neural network (ANN), the work in [34] used amplitude and time from GPR A-Scan to detect the shape, material and depth of a buried object. Extracting a signal envelope, peak detection of envelope and depth of buried empty tube from A-Scan through analytic signal technique [6]. Gilmore et.al [7] extracted features using the Hu's seven invariant moments algorithm, and latter passed them through an ANN classifier [29] to detect targets, however many false negatives were observed. While these techniques have been modestly effective, their performance is limited by the insufficient amount of real-world labeled GPR datasets for training the corresponding models or classifiers. To deal with the scarcity of GPR data, simulation-based methods have been proposed to increase the availability of training data, but these methods fail to represent the full spectrum of features found in real GPR data. Therefore, classifiers trained on simulated B-scan images tend to perform poorly on real world B-scan images. A successful remedy is the combination of simulated and collected real-world GPR data in training [25].

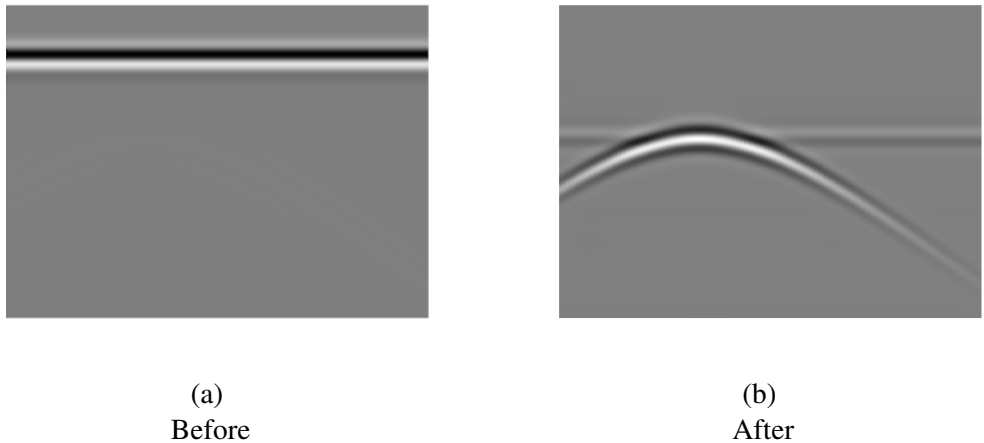

(a)  
Before

(b)  
After

Figure 2.3

Before and After Pre-processing

### *2.2.3 Frequency B-scan*

Figure 2.4

Frequency B-scan Creation Process

Figure 2.5

Frequency B-scan

## 2.3 Neural Networks

Neural Networks are a subset of machine learning algorithms that are inspired by neuron connections in the human brain. At the most basic level, the network consists of a single computational node that takes a vectored input, which is multiplied by a hidden weight, and transformed with a non-linearity to achieve an activated output. The idea is that the hidden weight can be optimized to produce a desired output. Networks may contain multiple nodes with many hidden layers. Some of the largest models can have hundreds of layers. For the purposes of this thesis, it is important to think of a neural network as an estimator of a probability distribution when given a conditional input. In fact, the activated output is commonly referred to as the posterior distribution. The name posterior simply indicates that it is the output distribution in respect of the prior, or to put succinctly, the inputs [9]. This probabilistic approach will be very important in the next section, as it is a vital attribute of understanding how to measure the difference between distributions which is a core principle of generative models.

## 2.4 Statistical Distance

Statistical Distance is simply a measure of the difference between two distributions. In the context of machine learning, it is possible to train a model to reduce this difference. In the following subsections, we will discuss several ways to measure this difference, and how we can apply neural networks as a means of reducing this difference [9].

### 2.4.1 Mean Squared Error

Although not technically a statistical distance metric, we can think of Mean Squared Error (MSE) as a means of calculating how different the expected output is from the observed. Equation 2.1 is a generalized form of this concept. If we recall from the previous section, that a neural network can output a conditional distribution. From this output, we can calculate the difference between itself and the expected output. Furthermore, we can optimize the model to minimize this difference to coerce the output to be more similar to our expected output. In the next part, we will discuss another method of measuring difference between the observed and expected. A measure that is specifically designed for this problem [9].

$$MSE(\hat{\theta}) = \mathrm{E}_{\theta}\left[(\hat{\theta} - \theta)^2\right] \tag{2.1}$$

*2.4.2 KL Divergence*

Kullback-Leiber divergence (KLD)[20] can be thought of as a measure of the difference between two distributions *P* and *Q*. Due to its asymmetric nature, KLD is not actually a true measurement of distance. However, the former conceptualization can remain for our purposes. An important feature of KLD is that it cannot be negative. Therefore, the minimum value is 0, at which, indicates that *P* and *Q* are the same distribution . Equation 2.2, depicts the mathematical formula of KLD. Similar to MSE, this difference can be minimized by a neural network and has been successfully used as a loss function [9]. However, the next measure is the most important in relation to this thesis.

$$D_{\mathrm{KL}}(P \parallel Q) = \sum_{x \in \mathscr{X}} P(x) \log\left(\frac{P(x)}{Q(x)}\right) \tag{2.2}$$

*2.4.3 Wasserstein Distance*

A distribution can be described in terms of probability mass. If we know the shape of the mass, we can derive a method that moves mass in one distribution to match a target distribution. This is the idea of Optimal Transport Theory and the Wasserstein metric is a solution to this problem. Commonly referred to as the earth mover's distance, the Wasserstein metric tells us how much mass we need to move to turn a given distribution into the target distribution. The task then becomes an issue of finding the proper function that transforms $p_r$ into $p_\Theta$ as shown in Equation 2.3. The importance in this, is that we can use a neural network to approximate the function, by training a model to minimize the Wasserstein distance. In approximating this function, we are able to take a prior distribution and transform it to closely match the target distribution. In the next section, we will see how this process can be applied to generate realistic data [28].

$$W(p_r, p_\theta) = \sup_{\|f\|_L \leq 1} \mathbb{E}_{s \sim p_r}[f(s)] - \mathbb{E}_{t \sim p_\theta}[f(t)] \tag{2.3}$$

## 2.5 Generative Models

Generative models are a subset of Neural Networks that enable the synthesis of realistic data. Research into this type of model has exploded over the last few years, mainly due to the types of problems that are able to be solved by them. For instance, generative models have been widely successful at generating realistic Speech, Music, Images, and Videos. The basis of this ability lies in the way neural networks are universal function approximators [13]. This allows a model to be fed inputs, learn the features of these inputs, and then produce outputs with the likeness of the input data. If we look at our inputs and outputs as probability distributions one can begin to see how it would be very important to be able to measure the differences in two probability distributions [9].

## 2.6 Generative Adversarial Networks and Applications

A Generative Adversarial Network (GAN) is a form of generative model, in which two separate models are entangled in a zero-sum game. The generative portion of the network is denoted as the generator ($G$). The goal of the generator is to synthesize the most realistic posterior distribution. In opposition of the generator, the discriminator ($D$), decides if the posterior distribution is legitimate, or a counterfeit. This process is carried out in tandem during training and can lead to instability [8]. The input to the generator is a Gaussian noise vector $\mathcal{N}(0,1)$. A transformation is applied to this vector, thus producing a posterior with equal dimensionality as the target distribution. Furthermore, this process is carried out by interpolation of the input vector through one or more deconvolution operations[citation]. The discriminator decomposes this output distribution into a binary probability through a sigmoid activation. The basics of this min-max game have changed very little since first proposal.

Generative Adversarial Networks have received wide attention in the machine learning field for their potential to learn high-dimensional, complex real data distribution [8, 12]. Specifically, they do not rely on any assumptions about the distribution and can generate real-like samples from latent space in a simple manner. This powerful property leads GANs to be used in many generative tasks to replicate the real-world rich content such as images, videos, speech, written language, and music [12]. There has been some work done on employing GANs for data augmentation in image classification using deep learning [12]. Furthermore, GAN can also be interpreted to measure the discrepancy between the generated data distribution and the real data distribution and then learn to reduce it. The discriminator is used to implicitly measure the discrepancy. Despite the advantage and theoretical support of GAN, many shortcomings have been found due to the practical issues and inability to implement the assumption in theory including the infinite capacity of the discriminator. There have been many attempts to solve these issues by changing the objective function, the architecture, etc. Moreover, the most recent additions to the adversarial framework have improved on many weak points in the original architecture. Wasserstein Loss has been used in GAN models to improve the stability of the adversarial game [4]. Moreover, this architecture can be further improved with the use of a gradient penalty term [10].

## 2.7 Evaluation of Generative Models

Many generative architectures use Mean Opinion Score (MOS) or other qualitative metrics for model evaluation [15]. This type of evaluation is readily available via Amazon Turkers or similar service that allows the general public to give an opinion. In our case, qualitative evaluation is unrealistic due to the requirement of domain expertise to detect the realistic nature of each synthesized GPR signal. Moreover, we would like to stray away from qualitative evaluation and use a more quantitative method. In this work, we validate the quality of generated output by an improvement factor in the recognitive ability of our object identification model. It is important to note the parallels between this technique and the commonly used Inception Score [27]. However, with the lack of widespread availability of benchmark classifiers applied to this domain, we use another method.

*2.7.1  Optimization Paradox*

Evaluation of a generative models presents a very unique challenge in respect to other machine learning models. As we recently covered, the idea is that we train a generative model to match the distribution of the training set in the output distribution. In minimizing this distance, we approach the distribution of the training set. The challenge is that at some point during this process variation is lost. Traditionally, this would be called over-fitting. However, detecting over-fitting in a generative model presents an evaluation paradox. The main question is: how do you maximize the difference in distributions as a metric for detecting over-fitting, when the entire principle of optimization is to reduce this difference? As one can see, this presents a rather tricky environment for evaluating generative models. The answer to the perfect evaluation method for generative models is a hot topic, that at the time of this writing, does not have a definitive solution. However, in principle there exists an optimization point which retains qualitative realism, but also allows for maximum variability. It is finding this balance, that encompasses all major challenges of generative model evaluation.

## 2.8  Object Recognition

Object Recognition can be simply thought of as taking an image, and determining what object is in that image. In the case of this thesis, the objects are the different materials of underground cylinders. This is essentially a classification problem. One takes images with $N$ classes and trains a model to predict correctly each of these classes when presented with a new image. An important note is that these types of models are only able to predict supervised classes. For instance, if a model was trained to be a dog detector, and only saw images of dogs. The model would perform poorly on images of a zebra. Therefore, in the scope of this thesis, the only concern is for the model to predict a class out of the $N$ classes available during training. However, it would be possible to add an additional class $(N+1)$ to allow for a catch all that accepts anything that is not one of the supervised classes. This is beyond the scope of this work, because the only images being used are known to contain one of the classes of cylinders.

## 2.9   Classification Evaluation Metrics

As we saw in Section 2.7, evaluation is an important aspect of machine learning. Below, the most common evaluation metrics for classification are discussed. A precursor to this discussion, requires a few simple definitions. True Positive (TP), is the number of correct predictions. In contrast, True Negative (TN) is the number of correct predictions for the negative class. Furthermore, a False Negative (FN) is when the class was actually negative, but the model predicted the class to be positive. Likewise, a False Negative (FN), is a positive class that is classified as negative.

### *2.9.1   Accuracy*

By far, the most ubiquitous evaluation metric is accuracy. As depicted in 2.4, accuracy is simply the fraction of true positives and true negatives over all of the observations. In Chapter 4., accuracy is reported. However, often it is more important to measure not only what the classifier gets right, but *how* wrong the classifier is. The following equations will shed light on the incorrectness of the model and allow us to make inferences based on this information.

$$\frac{TP+TN}{TP+TN+FP+FN} \tag{2.4}$$

### *2.9.2   Precision*

Precision is a measure of how often the model makes the correct prediction when looking at both correct predictions and predictions the model got wrong, when it was actually correct. Equation 2.5 is the equation for precision. It is the measure of True Positives over the total number of positive predictions.

$$\frac{TP}{TP+FN} \tag{2.5}$$

*2.9.3 Recall*

The next metric discussed is recall. To put plainly, this metric determines how often the model is right when giving a positive class prediction. Equation 2.6 shows that recall is defined as the number of True Positives over the total number of positive predictions. Now an even better evaluation metric would be the combination of precision and recall. F1-Score is exactly this metric.

$$\frac{TP}{TP+FP} \tag{2.6}$$

*2.9.4 F1-Score*

F1-Score combines both precision and recall into a single metric. This allows for ease of optimization. When looking at maximizing only the F1-Score, by default, it is also possible to maximize precision and recall. This is arguably one of the best metrics for classification problems. From Equation 2.7, it can be seen that F1-Score combines precision and recall to form a powerful evaluation metric.

$$2 \cdot \frac{precision \cdot recall}{precision + recall} \tag{2.7}$$

# CHAPTER 3
# METHODOLOGY

## 3.1 Introduction

The following sections are a detailed overview of the experiments conducted. First, training synthesis through gprMax is discussed. The topics covered are automation of training data generation and pre-processing. Then, a step by step look at the model architectures used, along with their respective loss functions, and the algorithm for training these models.

## 3.2 Software

For programming, python is the only language used. All machine learning experiments were carried out with `tf.keras` which is part of the main `tensorflow` package [1]. In addtion, for preprocessing we use numpy which is part of the `scipy` ecosystem [16]. Finally, for visualization, we use `matplotlib` [14].

## 3.3 Generating Training Data with GprMax

To acquire training data for the GAN model, we use gprMax, an open source software to simulate electromagnetic wave propagation. It solves Maxwell's equations in 3D using the Finite Difference Time Domain (FDTD) method [30]. We generate cylinders with diverse dielectric properties in range of substrate mixtures. For purposes of sample diversity we focus on a range of cylinder diameters in Peplinksi soil [24], with a range of sand to clay ratio for each image. To provide additional randomness, we apply a seed value that is randomly selected and applied to each iteration of training data generation. Therefore, each image produced by gprMax is unique. A total of 150 A-scan traces comprise the B-scan of a single simulation as show in Figure 3.1.

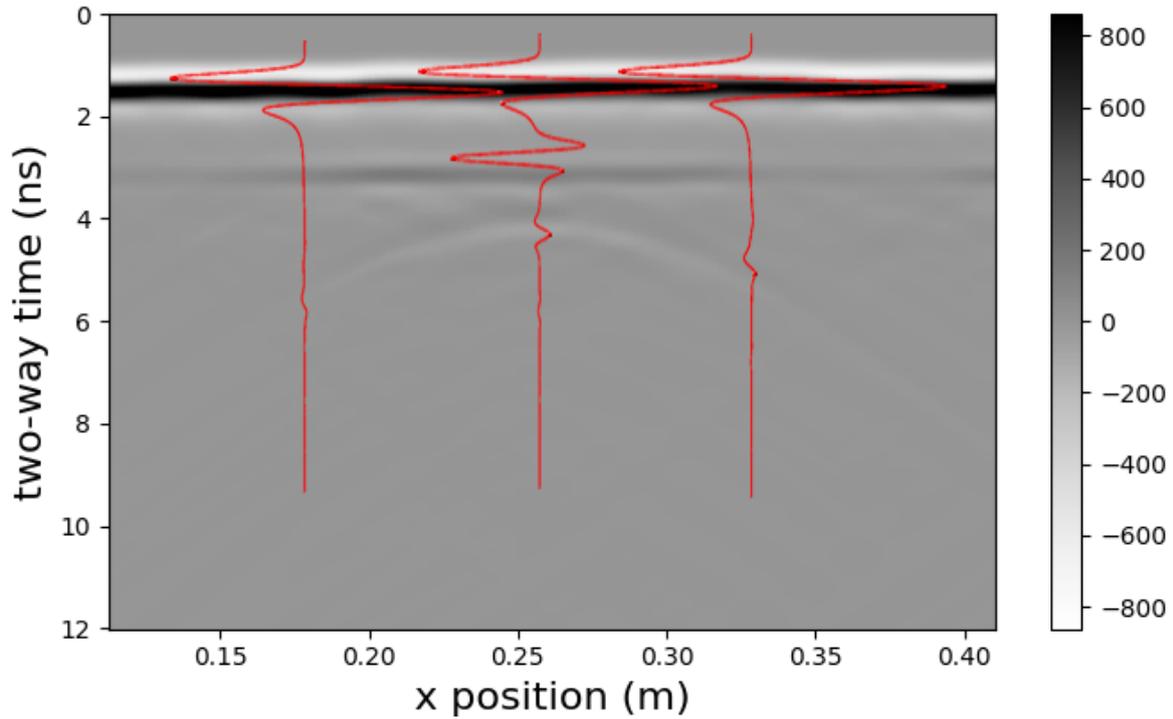

Figure 3.1

Generator and Discriminator Architecture

We re-size the B-scan to $(256, 256)$ with hamming interpolation. This centers the hyperbola, and ensures the the height and width dimensions are divisible by 2 for simplicity in upsampling $N(0, 1)$ in the generator and downsampling $G(z)$ in the discriminator. When producing the gprMax A-scans, we use dielectric properties that include three different pipe materials, metallic, plastic and concrete with radius ranging from 2 - 80mm, time window of 12e-9, and ricker waveform 1.5 GHz. The materials are used as classes, which allow us to condition the generator and identify the material with the object detection model.

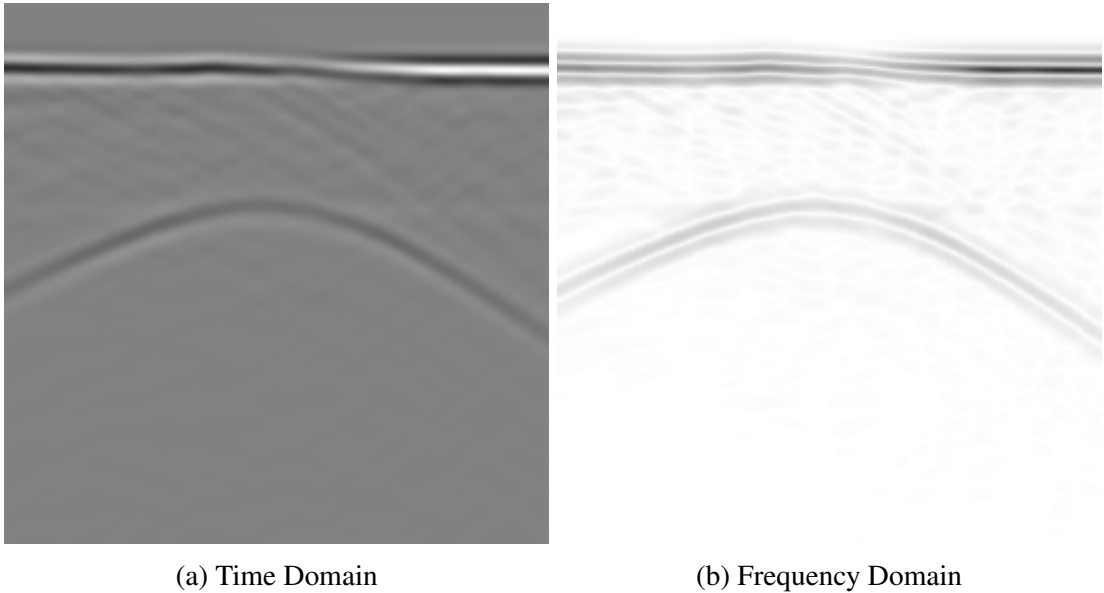

(a) Time Domain     (b) Frequency Domain

Figure 3.2

Time and Frequency Domain Training Data Example

## 3.4  Challenges with Using GprMax

First, it is necessary to mention that GprMax is a wonderful project and can do some really great things. However, using GprMax as a source for training data did not come without a set of challenges. The python interface is clunky and the documentation outlines only the most basic software features. The user is bound to a command line interface in python. This forces the user to sequentially generate each B-scan from the command line. Even with CUDA support, large models can take up to four hours on an Nvidia GTX 1080Ti. Now, imagine that the user wanted to generate one thousand B-scans. This would consume a great amount of the users time. An important note is that generation of many B-scans with varying features is possible. However, this task is left up to adding blocks of python in text files which has its flaws. For a solution to this we took a few steps to making this process user friendly. First, we managed to script the systematic of synthesis of B-scans with random sampling of feature combinations strictly in python without the use of the command line interface. Additionally, this can be accomplished in a Jupyter notebook though is not recommended for long generation sessions. Another solution was greatly reducing the model size. Unfortunately, this is a scaled down approach. Therefore, the models are not to the scale of real world scenarios. This relates to the aforementioned generation time. Fortunately, we were able to use the cluster at the SimCenter. In doing this, we were able to distribute the generation jobs across multiple GPU's and greatly reduce the time to generate a single B-scan to roughly three minutes. Figure 3.3 depicts a typical configuration file for GprMax.

```
#soil_peplinski: 0.5 0.5 2.0 2.66 0.001 0.25 my_soil
-----------------------------------------------------
#domain: 1.0 1.0 0.1
#dx_dy_dz: 0.002 0.002 0.002
#time_window: 12e-9
-----------------------------------------------------
#fractal_box: 0 0 0 1 0.75 0.1 1.5 1 1 1 50 my_soil my_soil_box 42 n
#add_surface_roughness: 0 0 0.1 0.1 0.1 0.1 1.5 1 1 0.065 0.080 my_soil_box
#cylinder: 0.33 0.33 0 0.33 0.33 0.002 0.01 pec y
-----------------------------------------------------
-----------------------------------------------------
#rx: 0.1125 0.1525 0
#src_steps: 0.002 0.0 0
#rx_steps: 0.002 0.0 0
-----------------------------------------------------
#waveform: ricker 1 1.5e9 my_ricker
#hertzian_dipole: z 0.150 0.170 0 my_ricker
#messages: y
```

Figure 3.3

GprMax Config File Example

## 3.5  GAN Architecture

The architecture proposed is a deep convolutional structure. Previous work suggested that a convolution with the filter with dimension of (5,5) is superior to other options for the modeling of ground penetrating radar data [3]. Therefore, we set the kernel size of all convolutions to 5 by 5. The generator, is conditioned to upsample a noise vector into a class from a supervised label. This is accomplished by introducing a label embedding vector and concatenating it with the posterior of the generator [21]. From this output, we calculate Wasserstein loss with gradient penalty [10] against the true image. The discriminator is used to determine validity of the generated output by directly comparing the two images. We calculate the Wasserstein distance between both real and fake images. In addition, the discriminator is trained to produce a predicted label for the generated image. For this output, we calculate categorical cross entropy between the predicted label and the true label. To improve overall model quality, we apply a Frequency domain loss function to $G(z)$.x

*3.5.1 Generator Architecture*

Table 3.1 depicts the architecture of the generator. As depicted, the generator takes the inputs of a random noise vector and a label for the desired class. The label in then passed through an embedding layer which allows for multiplication with the noise tensor. This is the vital step for the introduction of class conditioning and leaves us with a single input for the remaining layers. Next, the combined input is passed through a dense layer which gives us the dimensionality to be able to reshape the tensor into the 3-Dimensional shape of an image. From this point, we begin the upsampling process. We derive the upsampling method from [5], which indicates that many upsampling layers are favorable. In addition, we pass the upsampled vector though a convolution layer. This allows us to retain only the important information that we upsampled. Each time the input passes though an upsampling layer it doubles in size. The output is then activated with ReLU [22] and then passed through a batch normalization layer for regularization. We continue this process until the generator input is the same size as our target image. Finally, a Tanh [31] activation is applied to restrict the output to the range (-1, 1). The important part of the generator is that we want to learn the transformation of a noise vector into an image. We use a gaussian noise vector because it contains the least amount of prior knowledge [9]. Therefore, the primary learning objective is not what the generator learns from the noise vector, but how we can exploit the functional approximation property of neural networks to transform the noise vector into an image.

Table 3.1

Generator Architecture

| Operation | Output Shape |
|---|---|
| Input $N(0,1)$ | (n, 100) |
| Label | (n, 1) |
| Embedding | (n, 1, 100) |
| Flatten | (n, 100) |
| Multiply | (n, 100) |
| Dense(8*8*128) | (n, 8192) |
| Reshape(8, 8, 128) | (n, 8, 8, 128) |
| BatchNormalization(momentum=0.8) | (n, 8, 8, 128) |
| ConvTranspose2D(filters=128, kernel-size=5) | (n, 16, 16, 128) |
| Conv2D(filters=256, kernel-size=5, strides=1) | (n, 16, 16, 256) |
| ReLU | (n, 16, 16, 256) |
| BatchNormalization(momentum=0.8) | (n, 16, 16, 256) |
| ConvTranspose2D(filters=256, kernel-size=5) | (n, 32, 32, 256) |
| Conv2D(filters=128, kernel-size=5, strides=1) | (n, 32, 32, 128) |
| ReLU | (n, 32, 32, 128) |
| BatchNormalization(momentum=0.8) | (n, 32, 32, 128) |
| ConvTranspose2D(filters=128, kernel-size=5) | (n, 64, 64, 128) |
| Conv2D(filters=64, kernel-size=5, strides=1) | (n, 64, 64, 64) |
| ReLU | (n, 64, 64, 64) |
| BatchNormalization(momentum=0.8) | (n, 64, 64, 64) |
| ConvTranspose2D(filters=64, kernel-size=5) | (n, 128, 128, 64) |
| Conv2D(filters=32, kernel-size=5, strides=1) | (n, 128, 128, 32) |
| ReLU | (n, 128, 128, 32) |
| BatchNormalization(momentum=0.8) | (n, 128, 128, 32) |
| ConvTranspose2D(filters=32, kernel-size=5) | (n, 256, 256, 32) |
| Conv2D(filters=16, kernel-size=5, strides=1) | (n, 256, 256, 16) |
| ReLU | (n, 256, 256, 16) |
| BatchNormalization(momentum=0.8) | (n, 256, 256, 16) |
| Conv2D(filters=1, kernel-size=5, strides=1) | (n, 256, 256, 1) |
| Tanh | (n, 256, 256, 1) |

### 3.5.2 *Discriminator Architecture*

Table 3.2, is an example of the discriminator architecture. The discriminator accepts a tensor in the shape of a real image (256, 256, 1). During training, it receives both real and fake images and directly compares the two. This feedback is used to condition the generator to make better images. We use LeakyReLU[32] activation to reduce mode collapse, because the gradient after this activation is never 0. The downsampling pattern is a reversed version of the upsampling pattern in the generator. This adds additional balance to the training process which produces additional stability. Furthermore, we flatten the tensor before it enters the final Dense layer. This can be thought of as a summary of the information learned in the previous layers. Note, there is not a non-linearity applied to the output of the final layer. This is used for direct comparison with the real image.

Table 3.2

Discriminator Architecture

| Operation | Output Shape |
|---|---|
| Input | (n, 256, 256, 1) |
| Conv2D(filters=16, kernel-size=5, strides=2) | (n, 128, 128, 16) |
| LeakyReLU(alpha=0.3) | (n, 128, 128, 16) |
| Conv2D(filters=32, kernel-size=5, strides=2) | (n, 64, 64, 32) |
| LeakyReLU(alpha=0.3) | (n, 64, 64, 32) |
| Conv2D(filters=64, kernel-size=5, strides=2) | (n, 32, 32, 64) |
| LeakyReLU(alpha=0.3) | (n, 32, 32, 64) |
| Conv2D(filters=128, kernel-size=5, strides=2) | (n, 16, 16, 128) |
| LeakyReLU(alpha=0.3) | (n, 16, 16, 128) |
| Conv2D(filters=256, kernel-size=5, strides=2) | (n, 8, 8, 256) |
| LeakyReLU(alpha=0.3) | (n, 8, 8, 256) |
| Flatten | (n, 16384) |
| Dense(1) | (n, 1) |

### 3.5.3 Wasserstein Loss

To minimize the dissimilarity between *G(z)* and *y* we use Wasserstein Loss [4]. Likewise, this loss binds the discriminator. However, we include gradient penalty introduced in [10] to improve training stability.

$$W(p_r, p_g) = \max_{w \in W} \mathbb{E}_{x \sim p_r}[f_w(x)] - \mathbb{E}_{z \sim p_r(z)}[f_w(g_\theta(z))] \tag{3.1}$$

The final loss of the discriminator is defined by Equation (3.2) which includes the penalty term. This penalty term, proposed in [10], is an improvement on the gradient clipping outlined in [4].

$$L = \mathbb{E}\tilde{x} \sim \mathbb{P}_g[D_w(\tilde{x})] - [D_w(x)] + \lambda(\|\nabla_{\hat{x}} D_w(\hat{x})\|_2 - 1)^2 \tag{3.2}$$

### 3.5.4 Frequency Domain Loss

In Equation (3.3), $\phi$ represents the B-scan frequency transformation outlined in [33] with some minor modifications. Each A-scan that is contained in the B-scan has a Short Term Fourier Transform (STFT) [2] with 1024 FFT bins and 16 segment size applied. From these transforms, we take the max frequency from each A-scan. The max frequencies are then concatenated back together to form a Frequency B-scan. We do take the transformation a step further, by converting the Frequency B-scan to gray-scale. This allows for a 1:1 comparison with time domain B-scans. Furthermore, we are able to use the same number of channels in the model architecture.

$$\mathbb{E}_{(x,z) \sim \mathbb{P}}\left[\left||\phi(x) - \phi(G(z))|\right|\right] \tag{3.3}$$

### 3.5.5 Training Algorithm

Algorithm 1 is the process we follow for training the GAN model. This process does not differ greatly from [10]. However, we do make some modifications. The first modification is we adopt a faster $\alpha$ for Adam [18]. We found the original learning rate as specified in the paper to cause slower convergence. We also add 3.3 to line 9. This loss is added directly to the respective Wasserstein losses. We minimize the weighted sum of all losses for the generator and discriminator individually. The algorithm begins with samples of real data and latent variable. For each iteration of the training loop, the discriminator is updated a total of five times for each generator update. This is specified by the $n_{critic}$ argument, which is consistent in naming with the original paper.

**Algorithm 1** WGAN with gradient penalty.
We use default values of $\lambda = 10$, $n_{critic} = 5$, $\alpha = 0.0003$

**Require:** The gradient penalty coefficient $\lambda$, the number of critic iterations per generator $n_{critic}$, the batch size $m$, Adam hyper-parameter $\alpha$.
**Require:** initial critic parameters $w_0$, initial generator parameters $\theta_0$.

1: **while** validation loss is still decreasing **do**
2:    **for** t=1,...,$n_{critic}$ **do**
3:       **for** i=1,...,m **do**
4:          Sample B-scan $x \sim \mathbb{P}_r$, latent variable $z \sim p(z)$
5:          $\tilde{x} \leftarrow G_\theta(z)$
6:          $\hat{x} \leftarrow \varepsilon x + (1-\varepsilon)\tilde{x}$
7:          $L^{(i)} \leftarrow D_w(\tilde{x}) - D_w(x) + \lambda(\|\nabla_{\hat{x}} D_w(\hat{x})\|_2 - 1)^2$
8:          $L_2^{(i)} \leftarrow \left[ \|\phi(x) - \phi(G(z))\| \right]$
9:       **end for**
10:       $w \leftarrow Adam(\nabla_w \frac{1}{m} \sum_{i=1}^{m} L^{(i)}, w, \alpha)$
11:    **end for**
12:    Sample a batch of latent variable $\{z^{(i)}\}_{i=1}^m \sim p(z)$
13:    $\theta \leftarrow Adam(\nabla_w \frac{1}{m} \sum_{i=1}^{m} -D_w(G_\theta(z)), \theta, w, \alpha)$
14: **end while**

## 3.6 Object Identification Model

We train a separate auxiliary classifier to predict the object in the image. This is a basic classifier that uses cross entropy to create a separation boundary between classes. The architecture consists of two convolution layers that lead into a fully connected layer. The output of the final fully connected layer is activated with softmax to generate a categorical probability distribution. The loss function 3.4 is traditional categorical cross entropy.

$$-\sum_{c=1}^{M} y_{o,c} \log(\hat{y}) \quad (3.4)$$

We apply this loss to both the Time B-scan and Frequency B-scan to maximize the probability of a correct class prediction. Figure 3.3 depicts the basic classifier architecture. The basic classifier has two convolution layers, both activated with Leaky ReLU [32]. We use this opposed to traditional ReLU to mimic the architecture of the discriminator. In the initial tests we sought to use the discriminator as the classifier. However, this leads to extreme over fitting in the discriminator and poor performance for the classification task. Moreover, this also had a negative effect in the adversarial game, with the generator being able to constantly fool the discriminator. An important note in using a separate classifier is that this simple architecture can be a stand in for more complex object detection models such as Faster-RCNN[26] or Mask-RCNN[11]. This was an additional reason for not using the discriminator as the object detection model. To enable the use of the Time B-scan and Frequency B-scan, the architecture has a slight modification as depicted in Figure 3.4. The additional fully connected layer allows us to calculate a separate loss for the Frequency B-scan which is useful in training.

Table 3.3

Single Classifier Architecture

| Operation | Output Shape |
|---|---|
| Input B-scan | (n, 256, 256, 1) |
| Conv2D(filters=2, kernel size=1, strides=2) | (n, 128, 128, 2) |
| LeakyReLU(alpha=0.3) | (n, 128, 128, 2) |
| Conv2D(filters=4, kernel size=1, strides=2) | (n, 64, 64, 4) |
| LeakyReLU(alpha=0.3) | (n, 64, 64, 4) |
| Flatten | (n, 16384) |
| Dense | (n, 3) |

Table 3.4

Combined Classifier Architecture

| Operation | Output Shape |
|---|---|
| Input Time B-scan | (n, 256, 256, 1) |
| Input Frequency B-scan | (n, 256, 256, 1) |
| SingleClassifier(Time B-scan) | (n, 64, 64, 4) |
| SingleClassifier(Frequency B-scan) | (n, 64, 64, 4) |
| Multiply | (n, 64, 64, 4) |
| Flatten | (n, 16384) |
| Dense | (n, 3) |

The auxiliary classifier is trained in three scenarios. The first, data containing only images from gprMax are adopted. We use this performance as a baseline to compare our other experiments. The second, we use the full gprMax generated dataset with additional GAN generated images. Finally, we add a concurrent frequency domain optimization function to the generator, then train the object detection model to identify cylinder material from the b-scan with the assistance of information from the frequency representation.

# CHAPTER 4
# RESULTS AND DISCUSSION

## 4.1 Introduction

In the following sections, the results of the experiments along with their significance are presented and discussed. To begin, the GAN results are presented and discussed. Furthermore, many generated images are included for the reader to get a qualitative look. Finally, the object identification performance is displayed and discussed. This is the key area to identify the validity of the generative models ability to synthesize realistic images that improve the classification performance.

## 4.2 Unsupervised GAN Experiments with Real B-scans

In this section, we will be looking at experiments with real B-scan images. These are unsupervised because we do not have labels for these data. However, it is still possible to demonstrate that the GAN architecture can produce aesthetically pleasing images from field collected B-scans. These data were collected by University of Vermont at their GPR test site and consist of several underground cylinders of various material. An important distinction to make is that due to the lack of hyperbola variation in the training data, only certain hyperbola are possible in the generated sets. However, in GPR we actually want to limit the shape variation of the generated hyperbola. Therefore the focus is on variation in the noise surrounding the hyperbola. The next section demonstrates that our model is not limited in the capacity of background variation even with the retention of original hyperbola shape.

### *4.2.1 Realistic Noise Comparison*

Figure 4.1 demonstrates the level of noise variation in two similar B-scans. The hyperbola is virtually the same, but as one can see the noise surrounding has drastic differences. If we recall in Section 2.2, the noise is indicative of the dielectric properties of the surround soil. Therefore, a different level of background perturbation allows us to mimic different soil types that may have not been present in the training set. Resiliency to different soil types is a prime feature to have in a robust classifier for underground objects and a major focus of this work. Moreover, as will be demonstrated in Section 4.3, it is possible to condition the possible locations of the hyperbola if positional variation in the training set is present. In this case, there will be interpolation between all possible hyperbola positions. However, this still does not affect the shape or material of the hyperbola.

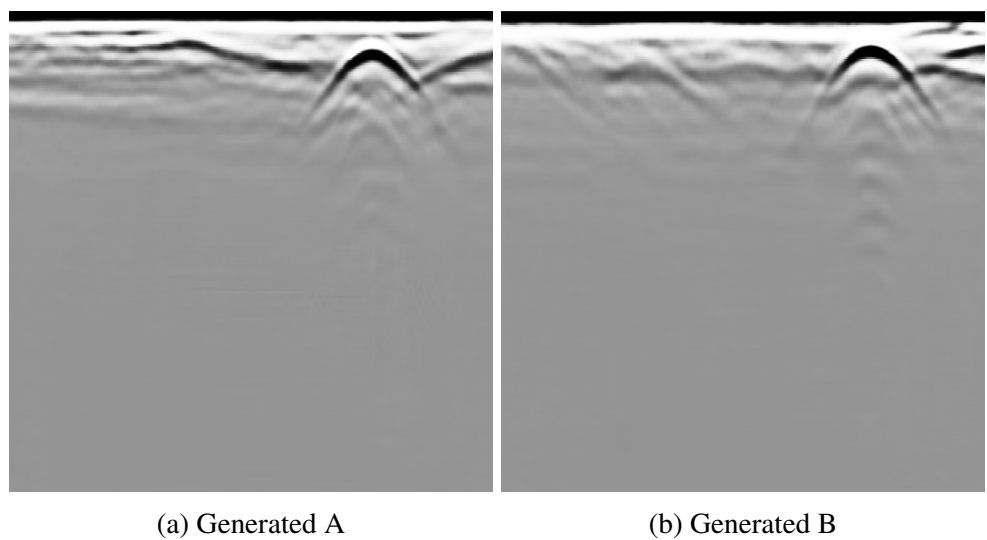

(a) Generated A    (b) Generated B

Figure 4.1

Soil Variance Comparison in Similar B-scans

### *4.2.2  Sample Diversity*

An important discussion of generative results is the diversity of the samples. In Figure 4.2, we can see that our generative model supplies a diverse set of samples. Note that the hyperbolas do not see much variation, this is the desired behavior. Moreover, one should pay special attention to how the noise changes between samples. Visually, not a single noise distribution is identical. In Section 2.7.1, we discussed a paradox when it comes to evaluating generative models. This paradox is especially important in sample diversity of unsupervised experiments which we do not have explicit class labels. The argument is that at what point are we simply memorizing the training distribution and how to define this point. Qualitatively, we see that our generative model does produce visually different images, but saying definitely that our model is not simply memorizing the training examples is a bit elusive. We do know that not a single identical image is produced, and that the images are visually realistic. However, exactly how close the images are to the training set is not quantitatively clear. The next section discusses the class conditioning results, which is a better approach to identify sample diversity.

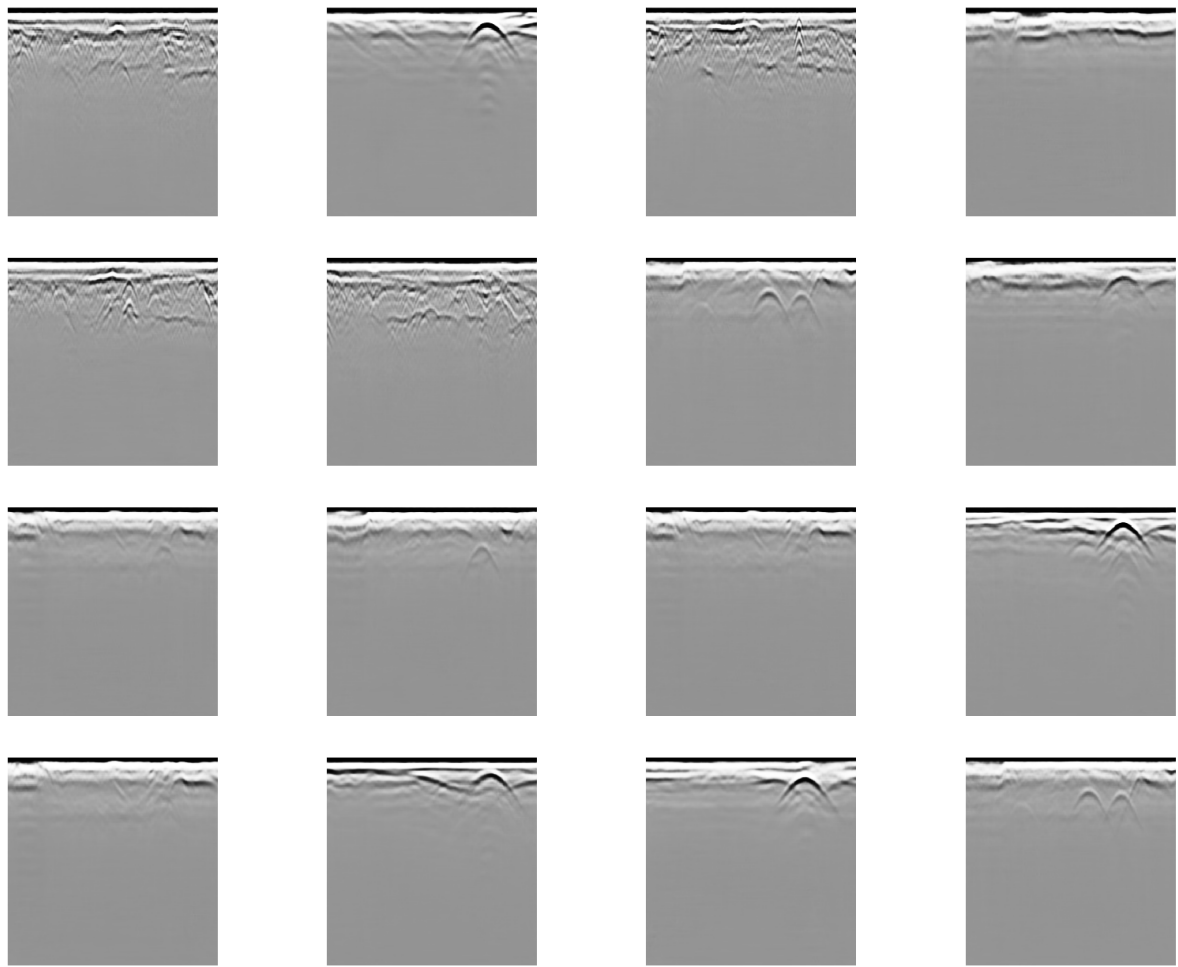

Figure 4.2

Realistic Generated Images

## 4.3 Supervised GAN Experiments

In this section, we will look at the results of Supervised GAN Experiments. These are the set of experiments that contain class conditioning. This is only possible with the use of GprMax that allows us to simulate the material of the underground object and retain a definitive label. This is necessary in the classification task and also a major drawback of collected B-scans. In the field, B-scans that are collected do not have a ground truth class label due to the subjective nature of real B-scan evaluation. In this experiment, we can generate realistic type data that does have a definitive class label. Therefore, we are able to map a set of image features to the label.

### *4.3.1 Class Conditioning Examples*

Figure 4.3 shows examples of the different classes that were the targets. The GAN model was able to learn distinct features of each class and then generate images of this type when given a label. An important note on continuation of this work is that ideally one would want to be able to combine the Unsupervised with the Supervised to generate a real B-scan with a known class. This is theoretically possible, however it is beyond the scope of this work.

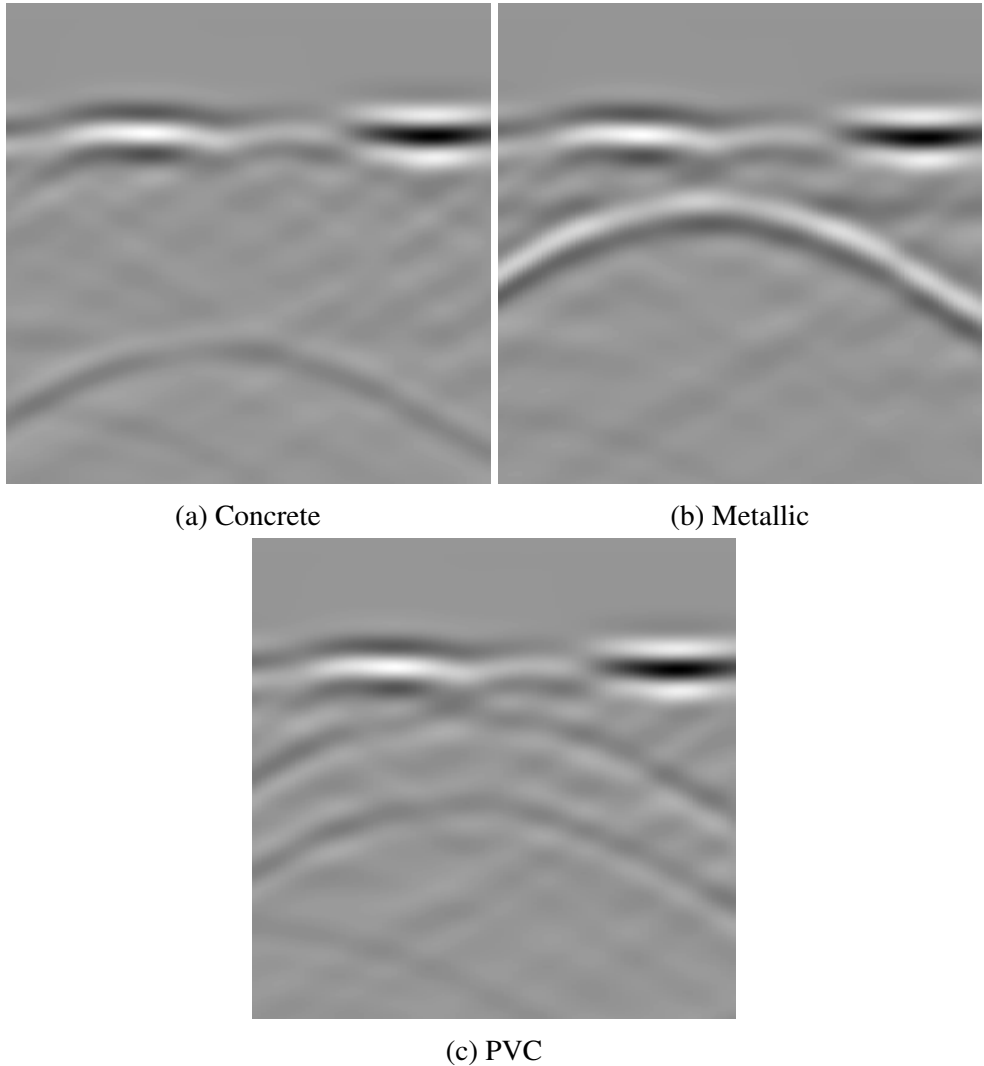

(a) Concrete  (b) Metallic

(c) PVC

Figure 4.3

Generated B-scan of Each Class

### *4.3.2 Sample Diversity*

To demonstrate the diversity in samples, the reader is directed to Figure 4.4. As one can see the diversity compared to the unsupervised experiments, especially in the hyperbola, is greater. This is due to the ability of being able to control the exact location of the hyperbola. Therefore, in the simulation through GprMax we can make certain that there is maximum variability in the training data. Furthermore, we get interpolation between positions of the hyperbola. This effect is due to defining a range where the hyperbola can occur. Moreover, with multiple hyperbola positions the model can learn the probable location of where a hyperbolas can occur which is evident by the high variability of hyperbola location in the generated set. Another important aspect is that even though the hyperbola changes position, it does not change shape and retains the characteristics of the particular material class. This is of vital importance because any change in shape or visibility can be interpreted as a different object than the desired target one.

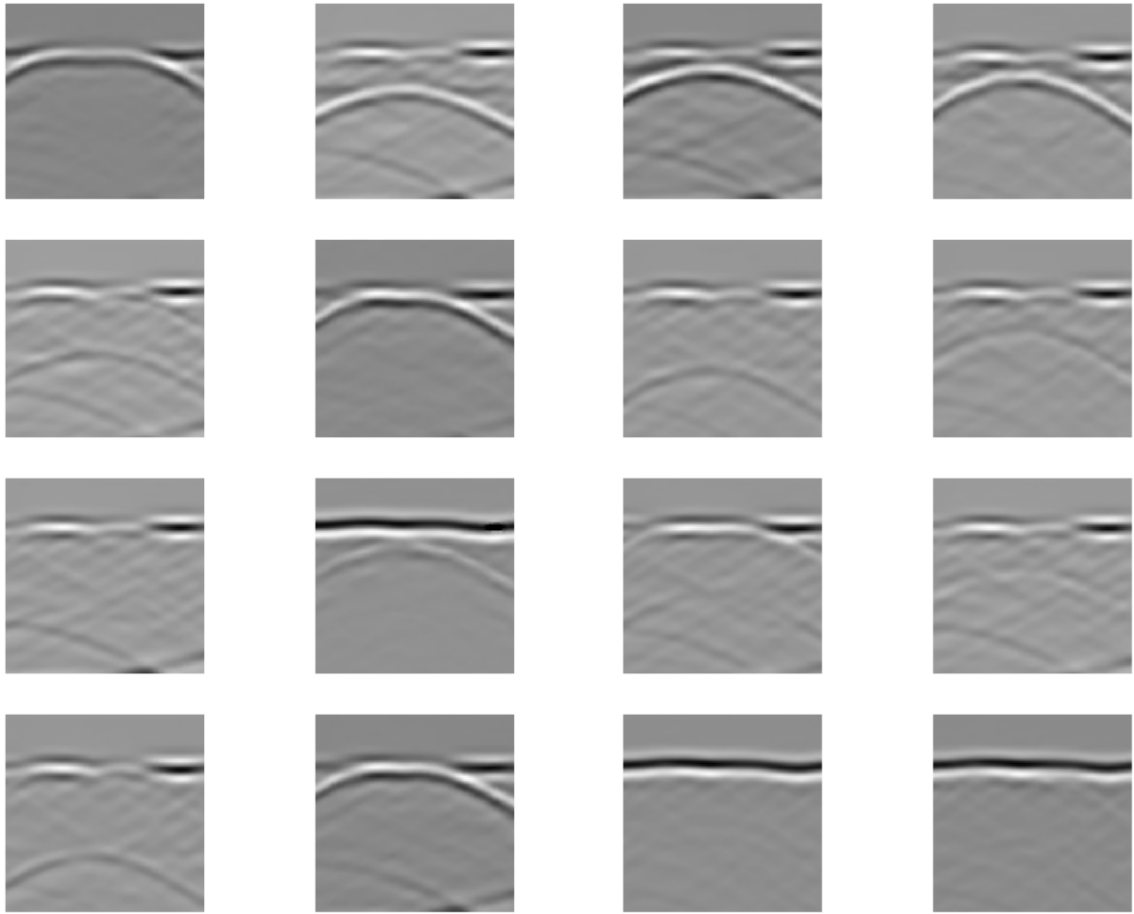

Figure 4.4

Generated Time B-scans

### 4.3.3 *Frequency Sample Diversity*

Figure 4.5 demonstrates the diversity in the generated Frequency B-scans. An important note is that these are not all of the possible variations, but a small subset for conciseness. In the frequency domain, we see that the model retains the ability to generate a wide array of features. Identical to the previous sections, we do not want to see variation in the hyperbola shape. Again, notice the noise variation, which ranges from very little noise in the background, to almost masking the hyperbola. This level of variation will be important in classifier evaluation.

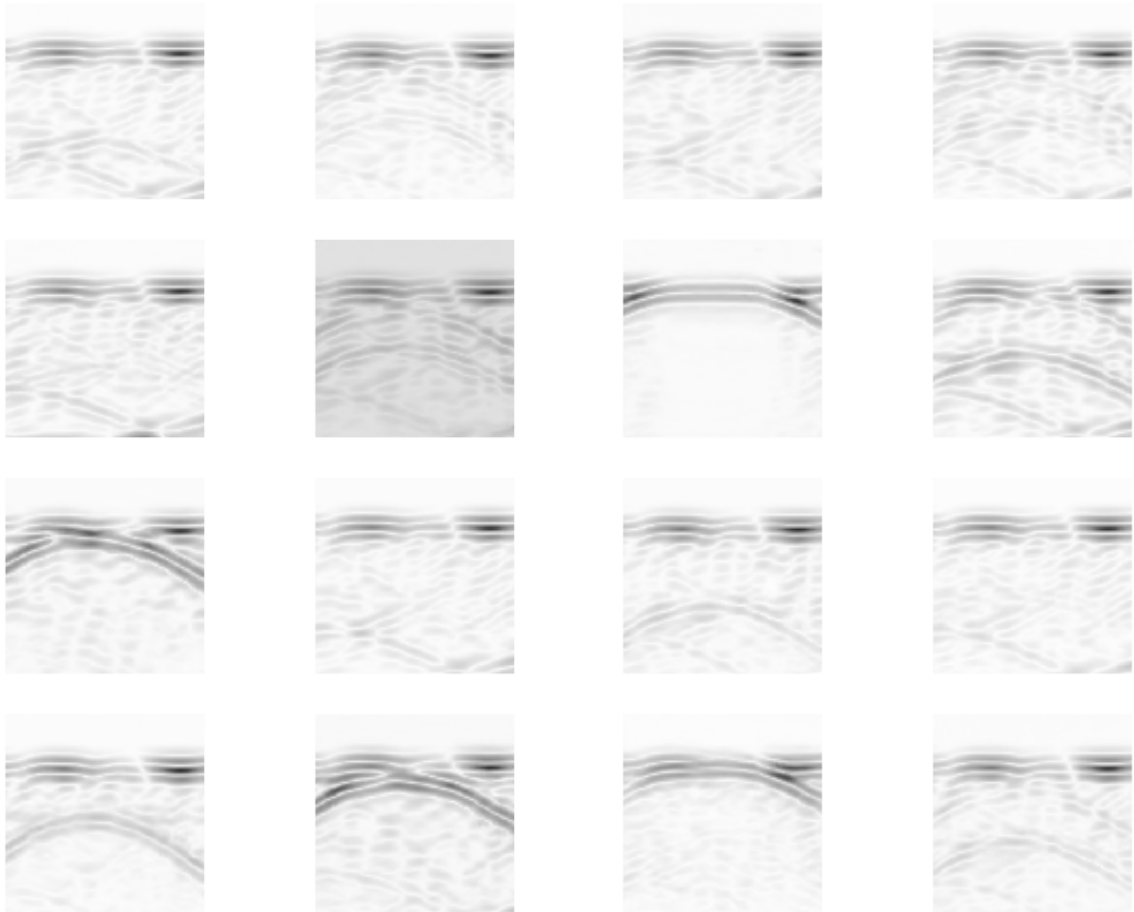

Figure 4.5

Generated Frequency B-scans using Figure 2.5

## 4.4 Object Identification Performance

The following sections are the performance of the proposed classifier in each test scenario. The main objective is to demonstrate improvement in two places. We would like to see performance improvement with data augmentation via GAN generated images and further improvement when combining the time and frequency B-scans.

*4.4.1 Time Domain*

For this set of experiments, we are only using time domain B-scans. These are the traditional B-scan representations that are present in the field collected examples presented in Section 4.2 and simulated in first part of Section 4.3. Performance in this area is a key baseline to realized performance in the subsequent sections.

Table 4.1

Baseline Object Detection Performance

| Before Augmentation | | | | | |
|---|---|---|---|---|---|
| Class | Accuracy | Precision | Recall | F1 | N |
| Concrete | 1.00 | 0.54 | 1.00 | 0.70 | 20 |
| Metallic | 0.81 | 1.00 | 0.82 | 0.90 | 11 |
| PVC | 0.50 | 0.67 | 0.57 | 0.62 | 14 |
| All | 0.77 | 0.74 | 0.80 | 0.75 | 45 |
| After Augmentation | | | | | |
| Class | Accuracy | Precision | Recall | F1 | N |
| Concrete | 1.00 | 0.88 | 1.00 | 0.93 | 14 |
| Metallic | 0.91 | 1.00 | 0.91 | 0.95 | 11 |
| PVC | 0.90 | 0.95 | 0.90 | 0.92 | 20 |
| All | 0.94 | 0.94 | 0.94 | 0.93 | 45 |

From Table 4.1, we present the tabular results of the baseline performance. It is important to point out that these results are still in the upper half in regards to the performance metrics. However, as it will be demonstrated, there is still room for improvement. PVC is by far the worst performing material class. This is due to the lack of reflectivity in PVC cylinders. In Figure 4.3, it can be seen that PVC is visually, the least prominent hyperbola followed closely by concrete. From the results, this visibility difference translates to the classifier performance. Metallic, being the most visually prominent, is easily identified by the detection model. The lower portion of Table 4.1 shows the performance achieved by training the classifier with augmented data. Overall, there is a performance increase when adding augmented training data. Most importantly, this is seen in the weak areas of the classifier. In the accuracy of PVC there is significant improvement that closes the gap between PVC and metallic cylinders. This means that when more samples are present in the training data that an overall increase in classifier performance will be realized.

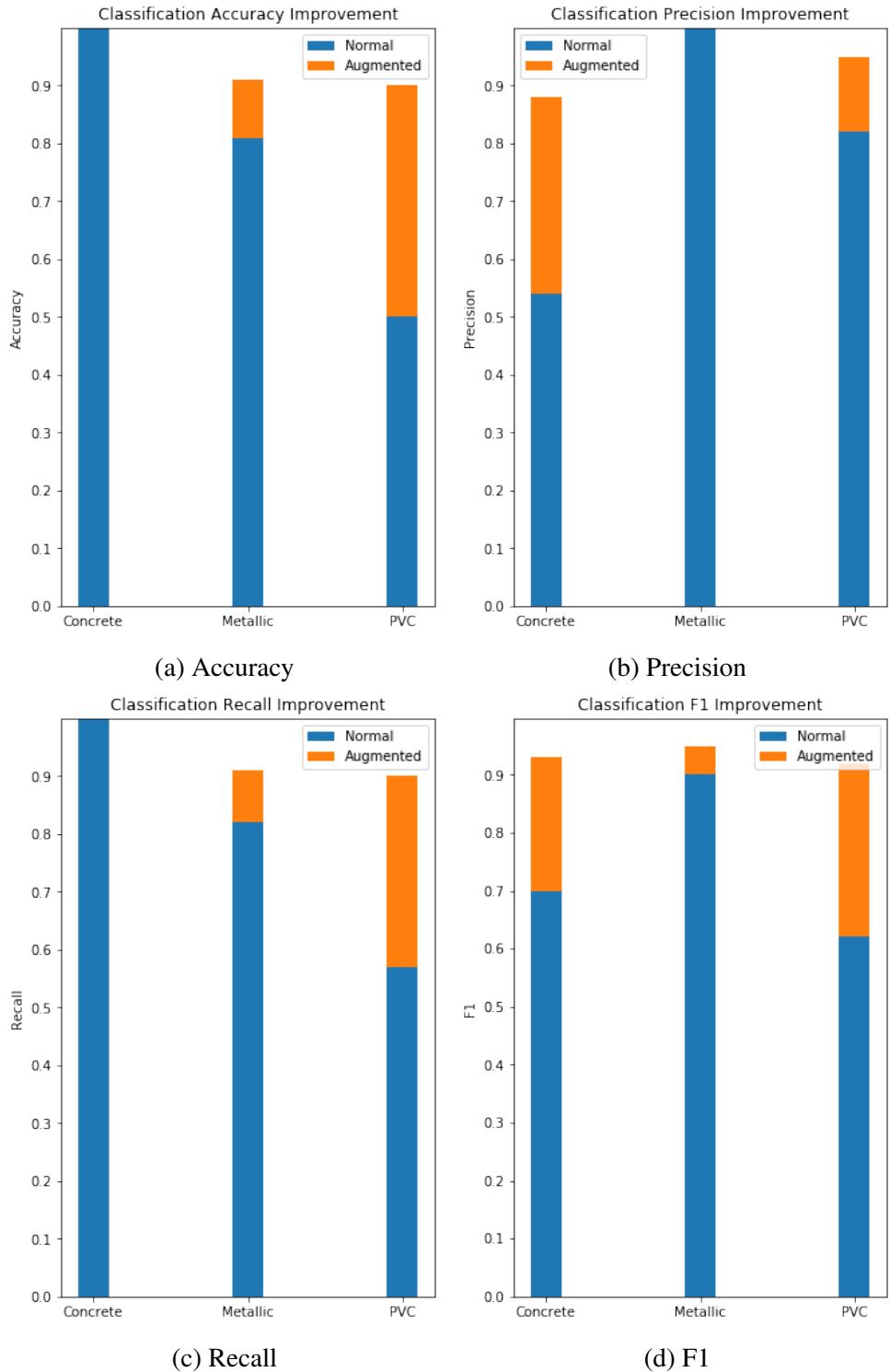

(a) Accuracy  (b) Precision

(c) Recall  (d) F1

Figure 4.6

Time Results

Figure 4.6 is a visual representation of the performance improvement. The blue bar is the performance without augmentation and the orange is improvement factor with augmentation. The total bar height is the final result achieved in each performance metric.

Note that for metallic cylinders performance does not have a large increase, but augmentation improved almost every metric for concrete and PVC cylinders. Next, we will see how the classifier performs in only the frequency domain.

### *4.4.2 Frequency Domain*

The next section discusses a classifier only trained on frequency B-scans. This experiment was to determine if the frequency representation leads to better performance in a particular class.

Table 4.2

Frequency Object Detection Performance

| Before Augmentation | | | | | |
|---|---|---|---|---|---|
| Class | Accuracy | Precision | Recall | F1 | N |
| Concrete | 0.21 | 0.80 | 0.21 | 0.33 | 19 |
| Metallic | 0.18 | 1.00 | 0.62 | 0.77 | 16 |
| PVC | 0.20 | 0.75 | 0.90 | 0.51 | 10 |
| All | 0.20 | 0.86 | 0.58 | 0.54 | 45 |
| After Augmentation | | | | | |
| Class | Accuracy | Precision | Recall | F1 | N |
| Concrete | 0.64 | 0.83 | 1.00 | 0.90 | 19 |
| Metallic | 0.36 | 1.00 | 0.88 | 0.93 | 16 |
| PVC | 0.60 | 1.00 | 0.60 | 0.67 | 10 |
| All | 0.53 | 0.94 | 0.83 | 0.83 | 45 |

Table 4.2 contains the numerical performance results. The classification of PVC outperformed that of metallic in accuracy when using a frequency B-scan. The significance in this is that if frequency information is given to the classifier that the weakest class in the baseline is able to be detected at a better rate than the strongest performing baseline class. Notice that precision in the frequency domain is high in all of the classes. Recall is an additional area in which PVC performs well. Although, performance is not quite as good as the time domain classifier trained with augmented data. Next, let us look at how augmentation can improve performance in the frequency domain. The bottom half of 4.2 contains the metrics after augmentation. Overall, there is improvement in all metrics.

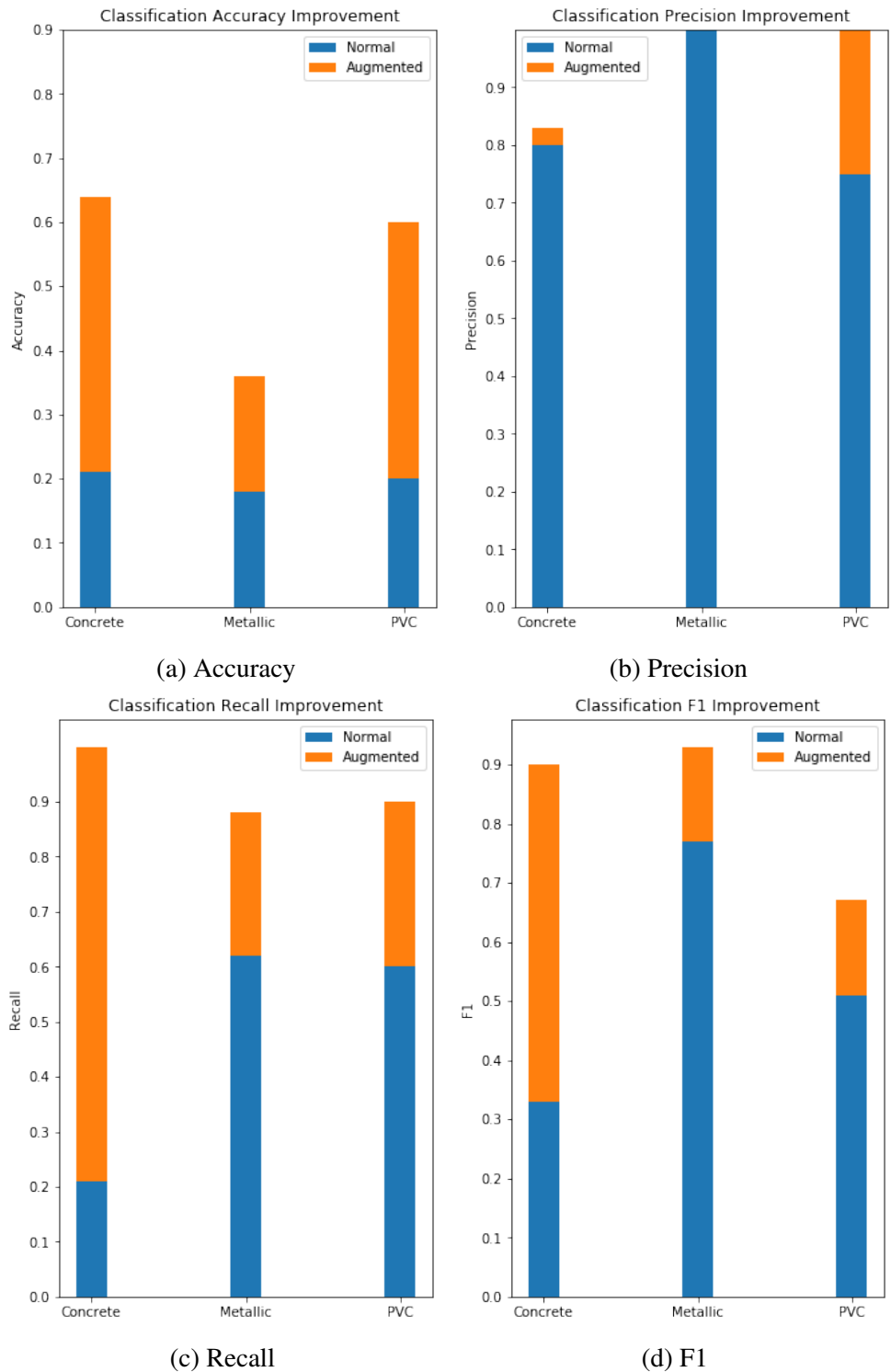

Figure 4.7

Frequency Results

Figure 4.7 is a visual demonstration of the improvement factor realized by using augmented data. An important area of improvement is in the Concrete class. Identification of Concrete was greatly improved by augmentation. Again, this is due to the improved detection ability that comes from using a frequency representation. This is evident when compared to the Metallic class. This class has the smallest improvement factor, because it was initially the best performing and also the most visible class. Finally, we will take a look at the performance when using a combined approach for material classification.

### 4.4.3 *Combined Approach*

In the combined approach, we are using both time domain and frequency domain B-scans. This essentially is doubling the amount of features used for classification. As reviewed in previous sections, frequency representations allowed improvement for the weak areas of material classification. A combined approach will yield an improved classifier for all materials.

Table 4.3

Combined Object Detection Performance

| Before Augmentation | | | | | |
|---|---|---|---|---|---|
| Class | Accuracy | Precision | Recall | F1 | N |
| Concrete | 1.00 | 0.54 | 1.00 | 0.70 | 18 |
| Metallic | 0.91 | 1.00 | 0.93 | 0.96 | 14 |
| PVC | 0.55 | 1.00 | 0.69 | 0.62 | 13 |
| All | 0.82 | 0.85 | 0.87 | 0.76 | 45 |
| After Augmentation | | | | | |
| Class | Accuracy | Precision | Recall | F1 | N |
| Concrete | 1.00 | 0.88 | 1.00 | 0.93 | 14 |
| Metallic | 1.00 | 1.00 | 1.00 | 1.00 | 11 |
| PVC | 0.95 | 1.00 | 0.90 | 0.95 | 20 |
| All | 0.98 | 0.96 | 0.97 | 0.96 | 45 |

Table 4.3, depicts the classification scores achieved before the use of augmented data. Compared to the baseline, this approach realizes a significant increase in all evaluation metrics. Notice that PVC is still the worst performer in accuracy. However, it still sees an improvement from the baseline. This indicates that using a combined approach did improve the results of a weak class in accuracy. This is also true for the other metrics in relation to PVC. Concrete did not see an improvement from the baseline when adding features from the frequency domain. This is unusual due to the increased performance in all metrics from the frequency domain experiments. However, it is important to note that Concrete already achieved max values in accuracy and precision in the baseline test. Therefore, the improvement did not occur in precision only. The two other classes, Metallic and PVC, saw performance improvement in every metric with the combined approach. Thus meaning that a combined approach is superior to using only time domain or frequency domain features individually. Now, we will look at the effect augmentation had on the combined classifier performance.

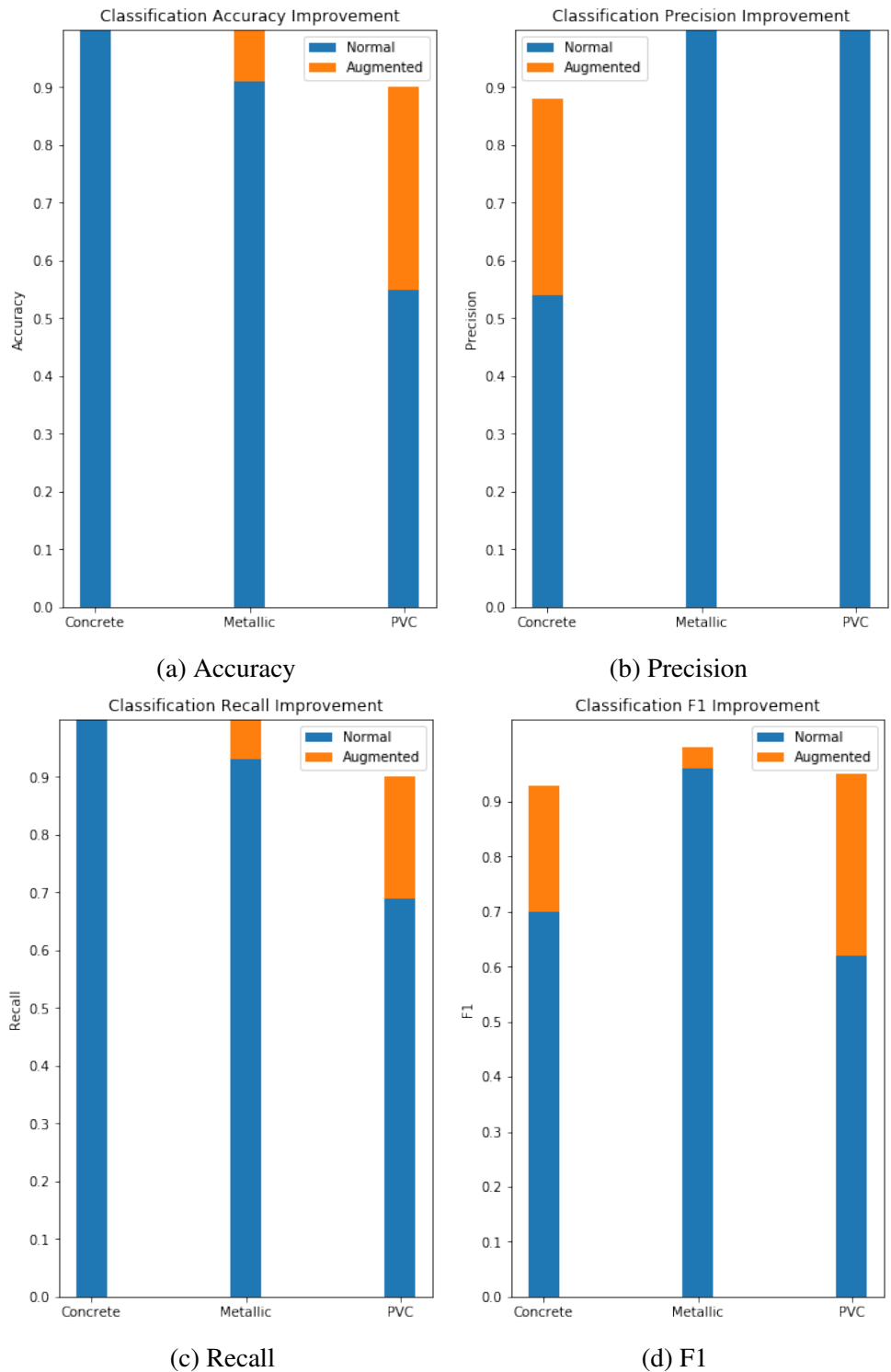

(a) Accuracy  (b) Precision

(c) Recall  (d) F1

Figure 4.8

Combined Results

Figure 4.8, shows the improvement factor of adding augmented data. The improvement factor is smallest among all of the experiments. This is due to the already

significant performance increase from incorporating frequency domain features. Continuing the previous trends, PVC is the weakest and also enjoys the most performance improvement from augmented data.

## 4.5 Summary

In conclusion, it was demonstrated that GAN could be successfully applied to GPR data. This is shown in both real and simulated B-scans. In addition, we looked at how the class conditioning could be applied to GAN to generate labeled training data for a classifier. With this training data, it was shown that GAN augmentation can improve a classifier. Furthermore, frequency domain features can be applied in a combined classifier, which enabled an additional boost in the scoring metrics for the classifier.

# CHAPTER 5

# CONCLUSION

## 5.1 Final Thoughts

The application of generative models to any domain, has its challenges. As we saw with applying GAN to GPR, there are some unique challenges to overcome. This thesis is a first step to further exploration of generative models for GPR. From this work, it was determined that GAN can successfully be applied to real GPR images. In addition, with the use of labeled training data, a conditional generative architecture can be applied for data augmentation. Furthermore, it was shown that a real-time classifier can be trained to detect the material of underground objects, and that this model can be improved with the incorporation of frequency domain features in classification. Moreover, With the addition of GAN synthesized data, we can train a classifier that detects objects with very high classification scores.

## 5.2 Future Work

Future research opportunities are, for the most part, infinite. However, the following are some immediate areas where improvement could be made as a continuation of this work.

### *5.2.1 Sophisticated Object Detection in Real Time*

The first major improvement would be the addition of a more complex object detection model. Due to the desire of real time constraints, a model similar to YOLO-Lite [23] would be ideal. Figure 5.1, shows some preliminary work in adding YOLO to GPR data. From the figure, it is evident that more work needs to be done in this area.

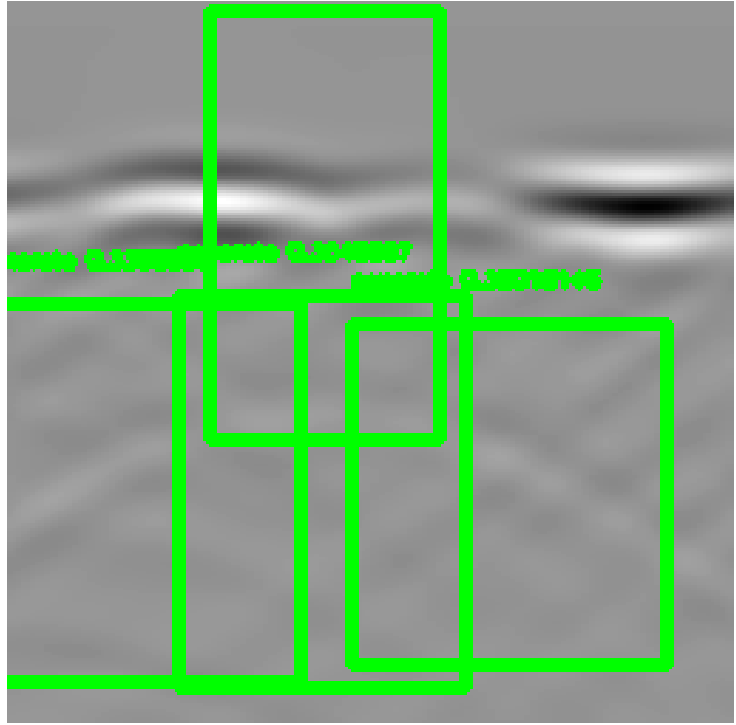

Figure 5.1

Preliminary Results With YOLO-Lite

This would not be the first work in the field on using sophisticated object detection. However, to my knowledge, an architecture that is capable of advanced object detection in real time has not been explored in the GPR space.

### 5.2.2  *Ideal GAN Architecture: gprGAN*

Recently, a major realization of the nature of GPR data came to mind. Ground penetrating radar data is sequential in nature. In review of earlier material, a single B-scan consists of multiple A-scans. Therefore, it follows that an architecture could exist that encapsulates the sequential aspect of GPR data. This is most important reason the proposed model architecture presented in this work, is not named gprGAN or similar, a trait commonplace in publishing work on adversarial models. A model capable of bearing the moniker gprGAN, would be able to synthesize sequences of A-scans to assemble a B-scan. Ideally, this model would be trained on one dimensional wave-forms, and have the ability to learn temporal dependencies in the segments of B-scans. Therefore, a coherent matrix of A-scans could be generated. The loss function would be a culmination of statistical distances between each A-scan in the sequence. Furthermore, it would be necessary to encompass the entire sequence of A-scans. Naturally, this sounds like a hierarchical model architecture and that could be one way of accomplishing this. In addition, it would be necessary to extending class conditioning to a multilabel problem. This would have diameter, material, and location learned independent of each other and the ability to mix these classes in conditional generation. However, a deeper exploration of this idea is necessary, but it would truly be a model capable of the title gprGAN.

# VITA

After graduating with a Bachelor of Science in Interdisciplinary Studies with a focus on Biology and Business from East Tennessee State University, Will Rice decided to change directions and pursue a Masters in Computer Science. In the first year of Masters study, he landed an internship in network administration at Life Care Centers of America. Within the first few months, he discovered a passion for machine learning which lead to an internship in speech and natural language processing at Pylon, ai where he was able to flourish in applications of machine learning and most importantly, generative models. Will assisted in the design and deployment of many deep learning models. The most complex being, a state of the art Text-to-Speech system which is a core component of the app RadioBrain. With a desire to enhance his technical writing, he decided that an important component to his education would be the completion of a thesis that applied generative models to a real world application. This was successfully accomplished with the guidance of Dr. Wu and Dr. Liang in the Networked Intelligence lab in the University of Tennessee at Chattanooga's Sim Center. In May 2019, he graduated with an MSc. in Computer Science.